\documentclass[sigconf]{acmart}

\makeatletter
\def\mdseries@tt{m}
\makeatother

\usepackage{color}

\usepackage{caption}
\usepackage[newfloat,draft=true]{minted}
\usepackage{times}
\usepackage{epsfig}
\usepackage{graphicx}
\usepackage{amsmath}
\usepackage{amsfonts}
\usepackage{amssymb}
\usepackage{calligra}
\usepackage{calrsfs}
\usepackage{mathtools}
\usepackage{hyperref}
\hypersetup{
    colorlinks=true,
    linkcolor=blue,
    filecolor=red,
    urlcolor=magenta,
    breaklinks=true,
}
\usepackage{breakurl}

\usepackage{booktabs}
\usepackage{multirow}
\usepackage{multicol}
\usepackage{siunitx}

\AtBeginEnvironment{minted}{%
  }

  \copyrightyear{2018}
  \acmYear{2018}
  \setcopyright{acmlicensed}
  \acmConference[KDD '18]{The 24th ACM SIGKDD International Conference on Knowledge Discovery \& Data Mining}{August 19--23, 2018}{London, United Kingdom}
  \acmBooktitle{KDD '18: The 24th ACM SIGKDD International Conference on Knowledge Discovery \& Data Mining, August 19--23, 2018, London, United Kingdom}
  \acmPrice{15.00}
  \acmDOI{10.1145/3219819.3219834}
  \acmISBN{978-1-4503-5552-0/18/08}

\newcommand{\eg}{\textit{e}.\textit{g}.~}

\begin{document}
\title{Corpus Conversion Service: \\A Machine Learning Platform to Ingest Documents at Scale.}

\author{Peter W J Staar, Michele Dolfi, Christoph Auer, Costas Bekas}
\email{ taa, dol, cau, bek @zurich.ibm.com}
\affiliation{%
  \institution{IBM Research}
  \streetaddress{14 Saumerstrasse}
  \city{Rueschlikon}
   \country{Switzerland}
}

\begin{abstract}
Over the past few decades, the amount of scientific articles and technical literature has increased exponentially in size. Consequently, there is a great need for systems that can ingest these documents at scale and make the contained knowledge discoverable. Unfortunately, both the format of these documents (e.g. the PDF format or bitmap images) as well as the presentation of the data (e.g. complex tables) make the extraction of qualitative and quantitive data extremely challenging. In this paper, we present a modular, cloud-based platform to ingest documents at scale. This platform, called the Corpus Conversion Service (CCS), implements a pipeline which allows users to parse and annotate documents (i.e. collect ground-truth), train machine-learning classification algorithms and ultimately convert any type of PDF or bitmap-documents to a structured content representation format. We will show that each of the modules is scalable due to an asynchronous microservice architecture and can therefore handle massive amounts of documents. Furthermore, we will show that our capability to gather ground-truth is accelerated by machine-learning algorithms by at least one order of magnitude. This allows us to both gather large amounts of ground-truth in very little time and obtain very good precision/recall metrics in the range of 99\% with regard to content conversion to structured output. The CCS platform is currently deployed on IBM internal infrastructure and serving more than 250 active users for knowledge-engineering project engagements.
\end{abstract}

%
%



\maketitle

\section{\label{sec:Introduction}Introduction}
It is estimated that there are roughly 2.5 trillion PDF documents currently in circulation\footnote{This number originates from a keynote talk by Phil Ydens, Adobe's VP Engineering for Document Cloud. A video of the presentation can be found here: \url{https://youtu.be/5Axw6OGPYHw} }. These documents range from manuals for appliances, annual reports of companies, all the way to research papers, detailing a specific scientific discovery. It is needless to say that valuable qualitative and quantitative information is contained in many of them. However, content encoded in PDF is by its nature reduced to streams of printing instructions purposed to faithfully present a pleasing  visual layout. Both the data representation and the enormous variability of layouts across these documents make it extremely challenging to access content and transform it into a representation that enables knowledge discovery.
In addition to the sheer current quantity of documents, the submission rate of published documents in the scientific domain is also growing exponentially\footnote{This is clearly the case on the popular arXiv scientific online repository:
\url{https://arxiv.org/help/stats/2012_by_area/index}}. This poses a real problem, since more and more information published in the PDF documents is going \textit{dark}.
In order to make the content of these documents searchable (e.g. \textit{find me a phase-diagram of material XYZ}), one needs essentially two components. First, you need to ingest documents from a variety of formats (with the PDF format being the most prevalent one) and convert these documents to structured data files with a structured format such as JSON or XML. Second, you need a query engine that is able to deal with a large variety of concepts (documents, images, authors, tables, etc) extracted from these documents and put these into context.

\begin{figure*}[th!]
\center
\includegraphics[scale=0.76]{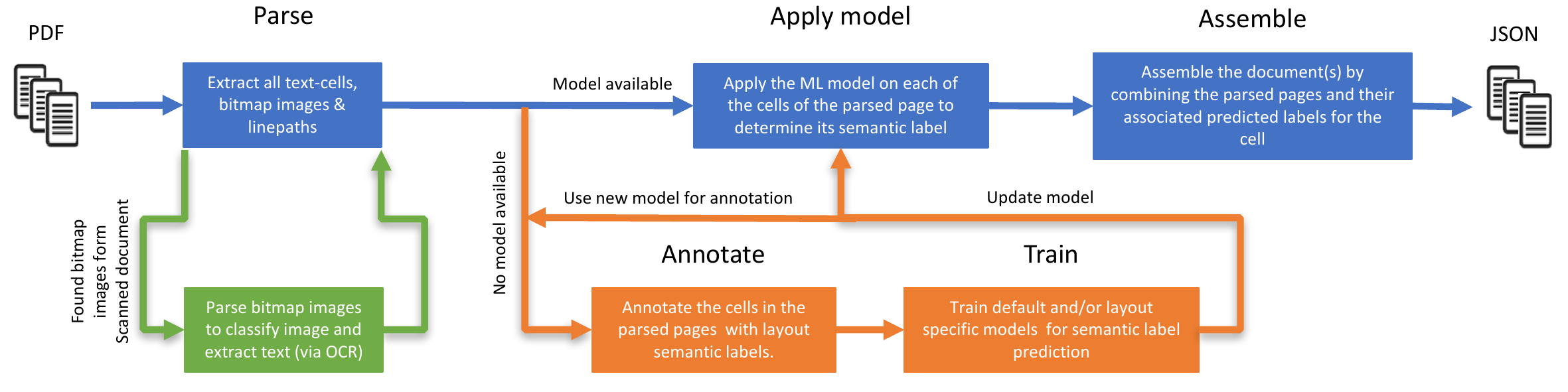}
\caption{\label{fig:SApipeline} A diagram of the conversion pipeline in the \textit{Corpus Conversion Service} platform. It consists of 5 components: (1) Parsing of the document and its contained bitmap images, (2) Annotating the text of the parsed documents with layout semantic labels, (3) Training models based on the ground-truth acquired by the annotations, (4) Applying machine learned models on the parsed documents to determine the layout semantic label of each cell and finally (5) Assembling the document into a structured data format (\eg JSON). The main conversion pipeline is depicted in blue and allows you to process and convert documents  at scale into a structured data format. The green and orange sections can be used optionally, in order to process scanned documents (green) or train new models based on human annotation (orange). }
\end{figure*}

In this paper, we focus entirely on the first component, the ingestion of documents and their conversion into structured data files. The solution we propose is thought of as a platform, which at its core has trainable machine learning algorithms.
This platform, called Corpus Conversion Service (CCS), consists out of a set of microservices organized in five main components.
Each of these microservices can be consumed by its own REST API. This approach not only allows us to build complex pipelines to process documents automatically, but also allows us to develop new microservices against the platform.
In order to make this platform scalable, all microservices are integrated through asynchronous communication protocols, which gives us many benefits: It allows to do proper resource management, eliminates strong dependencies and makes the platform robust against single task failures.

To obtain a thorough understanding of what our platform can do and how well it performs, we have structured this paper as follows: In Section~\ref{sec:SOTA}, we briefly review the current state-of-the-art document processing solutions. In Section~\ref{sec:Method}, we present the design of the platform and its components.
In Section~\ref{sec:Arch}, we discuss the architecture, the deployment methods, and how well the platform scales with regard to volume (both in users and content) and compute resources, respectively. Finally, in Section~\ref{sec:Conclusion}, we discuss the open questions w.r.t. research and possible next steps in the development of the platform.

\section{\label{sec:SOTA}State of the Art}

The task of converting PDF documents and automatic content reconstruction has been an outstanding problem for over three decades~\cite{Cattoni98geometriclayout, Chanod2005}.
Broadly speaking, there are two types of approaches to this problem. In the first approach, documents are converted with the goal to represent the content as close as possible to the original visual layout of the document. This can be done through a conversion from PDF towards HTML or MS Word for example. The second approach attempts to convert the document into a format that can be easily processed programmatically, i.e. a representation of the document which is not preserving the layout, yet contains all the content from the original document in a structured format. For example, this could be a JSON/XML file with a particular schema. Since our Corpus Conversion Service is thought of as a first step towards a knowledge discovery platform for documents, we have opted for the second approach in our solution.

Many solutions have already been developed that tackle the problem of document conversion. There are well known open-source programs such as Xpdf\footnote{\url{https://www.xpdfreader.com}} and Tabula\footnote{\url{http://tabula.technology/}}. There are also proprietary solutions, such as Abby\footnote{\url{https://www.abbyy.com/}}, Nuance\footnote{\url{https://www.nuance.com/}} or DataCap\footnote{\url{https://www.ibm.com/us-en/marketplace/data-capture-and-imaging}}. In contrast to the open-source solutions, all three proprietary solutions support also extraction from scanned documents. Besides the well known open-source and proprietary solutions, there are also countless academic solutions as well as libraries. For example, the challenge of segmenting complex page layouts is actively addressed by recurring competitions posed by ICDAR, as in Ref.~\cite{icdar-challenge-2015} and previous editions.

\section{\label{sec:Method}Platform design}

Given the plethora of existing solutions, we would like to point out how our solution differs from these, and thus approaches the problem of document conversion in a new way.

The key idea is that we do not write any rule-based conversion algorithms, but rather utilize generic machine learning algorithms which produce models that can be easily and quickly trained on ground-truth acquired via human annotation. This flexible mechanism allows us to adapt very quickly to certain templates of documents, achieve very accurate results and ultimately eliminates the time-consuming and costly tuning of traditional rule-based conversion algorithms. This approach is in stark contrast to the previously mentioned state of the art conversion systems, which are all rule-based.

While the approach of swapping rule based solutions with machine learning solutions might appear very natural in the current era of artificial intelligence, it has some serious consequences with regard to its design.
First of all, one can not think anymore at the level of a single document. Rather, one should think at the level of a collection of documents (or a corpus of documents). A machine learned model for a single document is not very useful, but a machine learned model for a certain type of documents (e.g. scientific articles, patents, regulations, contracts, etc.) obviously is. This is the first big distinction between the current existing solutions and ours: Existing solutions take one document at a time (no matter its origin) and convert it to a desired output format. Our solution can ingest an entire collection of documents and build machine learned models on top of that. Of course, once the the model is trained, one can convert documents one at a time, too.

A second discriminator between the existing solutions and ours is that we need to provide the tools to gather ground-truth, since no model can be trained without it.
Hence, not only do we need the ability to manage collections of documents, we also need the ability for people to annotate documents and store these annotations in an efficient way. These annotations are then used as ground-truth data to train models.
It is clear then that ML models add an extra level of complexity: One has to provide the ability to store a collection of documents, annotate these documents, store the annotations, train a model and ultimately apply this model on unseen documents. For the authors of this paper, it was therefore evident that our solution cannot be a monolithic application. It fits much better the concept of a cloud-based platform that can execute the previously mentioned tasks in an efficient and scalable way.

\subsection{\label{Components}Components}
Our platform implements a processing pipeline to ingest, manage, parse, annotate, train and eventually convert the data contained in any type of format (scanned or programmatically created PDF, bitmap images, Word documents, etc.) into a structured data format (e.g. JSON or XML).

\begin{figure}[!t]
\centering
\frame{\includegraphics[width=3.2in]{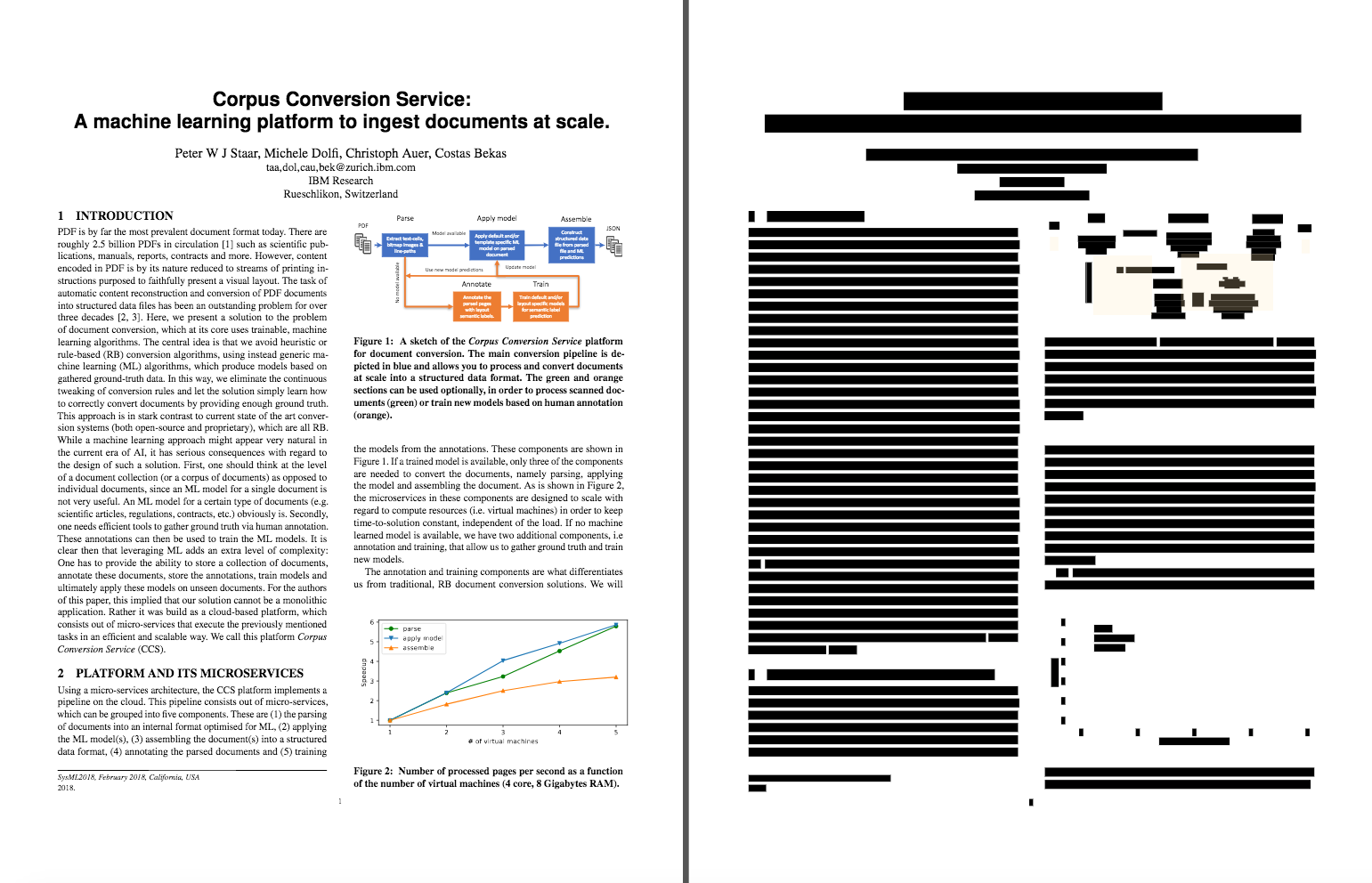}}
\caption{\label{fig:SAParsing} The cells obtained for the title page of a poster abstract about the CCS~\cite{ccs-poster} after the parsing stage. During the parsing, we extract all bounding boxes of the text (or \textit{cells}) in such a way that they all have: (1) a maximum width, (2) are only single line and (3) split into multiple cells in case of list-identifiers, multi-columns or crossing vertical lines (such as in tables).}
\end{figure}

This processing pipeline is formed by five components as depicted in Figure~\ref{fig:SApipeline}: (1) parsing of documents into an internal format optimised for ML, (2) Annotation of the label ground-truth in parsed documents (3) training ML models from the acquired annotations, (4) applying the custom ML model(s), (5) assembling the document(s) into a structured data format.
If a trained model is available, only components 1, 4 and 5 are needed to convert the documents. If no template-specific machine learned model is available yet, we provide two additional components 2 and 3, that allow users to gather ground-truth and train custom models. It is important to note that the platform comes with default models, so annotation and training are advised to retrieve the best quality output, yet they are optional.

Let us now elaborate on what each of the five components deliver in the rest of this section.
\subsection{\label{CCSParsing}Parsing of Documents}

\begin{figure}[!t]
\centering
\frame{\includegraphics[width=3.2in]{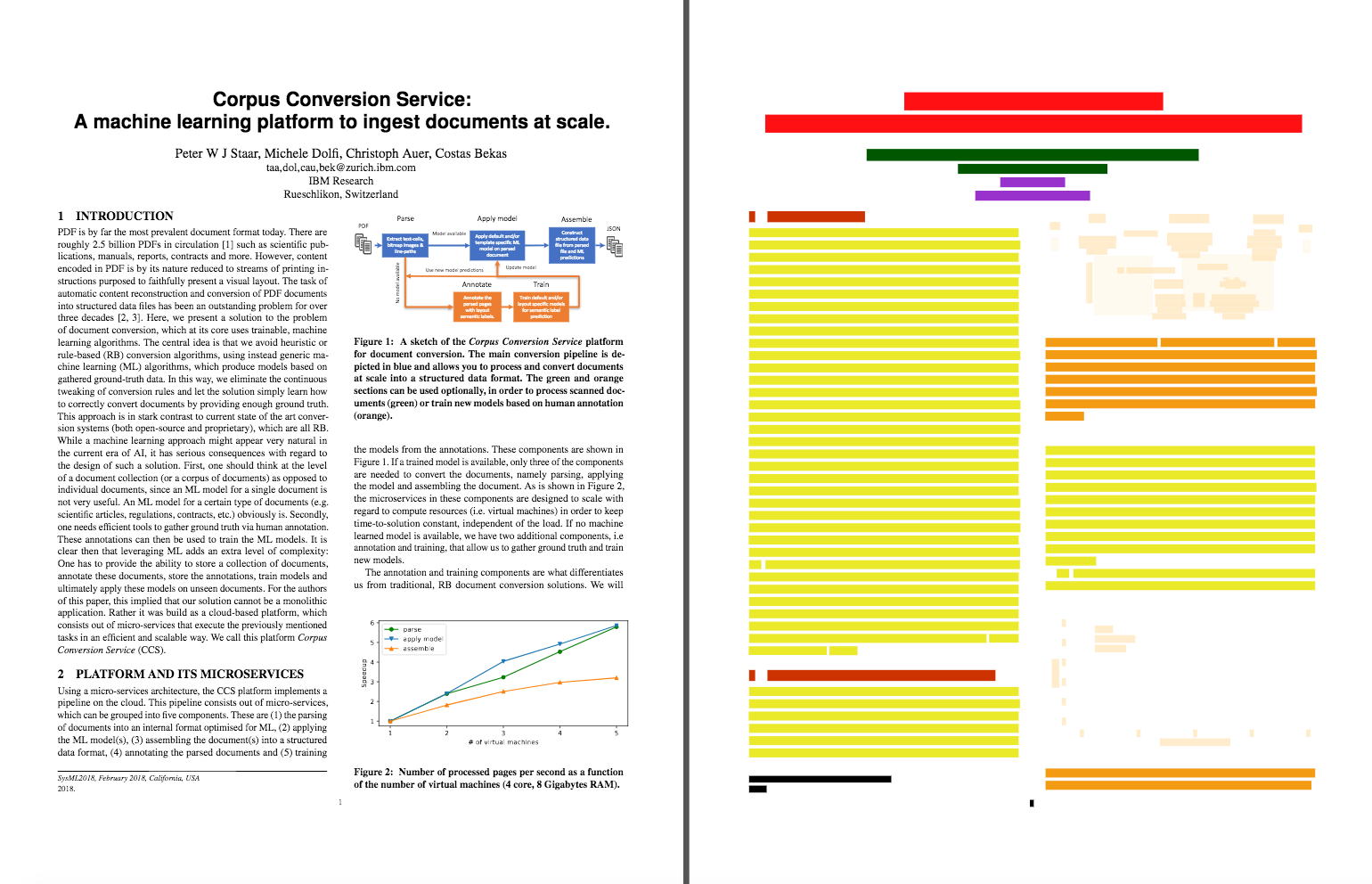}}
\caption{\label{fig:SAAnnotation} The labelled cells annotated on the title page of a poster abstract about the CCS~\cite{ccs-poster}. Here, the \textit{title}, \textit{authors}, \textit{affiliation}, \textit{subtitle}, \textit{main-text}, \textit{caption} and \textit{picture} labels are represented respectively as red, green, purple, dark-red, yellow, orange and ivory.}
\end{figure}

In the parsing component of the pipeline, we solve the following straightforward but non-trivial task: \emph{Find the bounding boxes of all text-snippets that appear on each PDF page.} For simplicity, we will refer to the bounding boxes of the text-snippets as \textit{cells} in the remainder of the paper. There are two reasons why we are interested in these cells. First, they provide us with the crucial geometric features which are later used in the machine learning models to determine the layout semantic label. Second, the concept of a cell can be easily transferred to scanned documents. In Figure~\ref{fig:SAParsing}, we show the cells obtained from an example PDF page after the parsing stage.

While the task of finding the cells might appear intuitive from a conceptual point of view, it is not in practice, since there does not exist a unique, precise definition of the cells. This lack of a precise definition has its origins not only in the ISO-standard\footnote{a line of text might be printed character-by-character, word-by-word or the entire text snippet.} detailing the PDF document code but also in the variability of the quality of PDFs. Older PDFs which were created from scanned images using OCR typically return cells for each word, while more recent PDFs allow us to create cells for full text-lines. This variability in the geometric features of the cell (e.g. the width of the cell) can negatively impact the performance of later machine learning models. As a consequence, we reduce the variability of the geometric features as much as possible. The more consistent and homogeneous the geometric features of a cell are, the better the machine learning algorithms can do predictions.

For programmatic PDFs, the text cells are contructed from raw streams of symbols and transforms defined in the PDF document. This operation relies on the iterators provided by the QPDF library\footnote{\url{http://qpdf.sourceforge.net/}}.

For scanned PDFs, we use a two step approach to find the cells by first running all bitmap resources in the PDF through an OCR engine and then merging the extracted text-snippets from the images with the remaining cells from the programmatically created content.
Eventually, all the created cells and line paths are stored in an internal JSON format, which also keeps references to the bitmap resources embedded in the PDF document. From this point, all further processing does not need to distinguish between scanned or programmatic sources.

\subsection{\label{ss:ha}Ground-truth gathering through human-annotation}
In this component, we collect ground-truth for the custom machine learning models to be trained on. Representative ground-truth data is of paramount importance to obtain machine learned models with excellent recall and precision. Unfortunately, it is often very hard to obtain lots of representative ground-truth data, primarily due the the enormous variability across the layout of documents. As a consequence, the concept of \emph{annotators} for documents were incorporated into the platform from the very beginning. The purpose of these annotators is two-fold.

First and foremost, the annotators on the platform allow us to gather ground-truth at scale using a crowd-sourcing approach. In each annotation task, we retrieve the original PDF page and its associated parsed components, containing the cells (see Figure~\ref{fig:SAParsing}). We then ask the (human) annotator to assign each cell a layout semantic label. Examples of semantic labels are: \textit{Title}, \textit{Abstract}, \textit{Authors}, \textit{Subtitle}, \textit{Text}, \textit{Table}, \textit{Figure}, \textit{List}, etc\footnote{It is important to notice that there is no restriction on the number of labels nor the semantic meaning of these labels. The only limitation one has is that the set of semantic labels needs to be consistent across the dataset, but this is evidently true for any type of ML algorithm.}. In the annotator tool, each layout semantic label is visually represented by a colour. By assigning a colour to each semantic label, the task of \textit{semantic} annotation is translated into a \textit{colouring-task}, as can be seen in Figure~\ref{fig:SAAnnotation}. Since humans are very efficient in visual recognition, this task comes very natural to us. The required time spent to annotate a single page starting from the parsing output has shown to average at 30 seconds over various annotation campaigns.

\begin{figure}[t]
\centering
\includegraphics[width=3.2in]{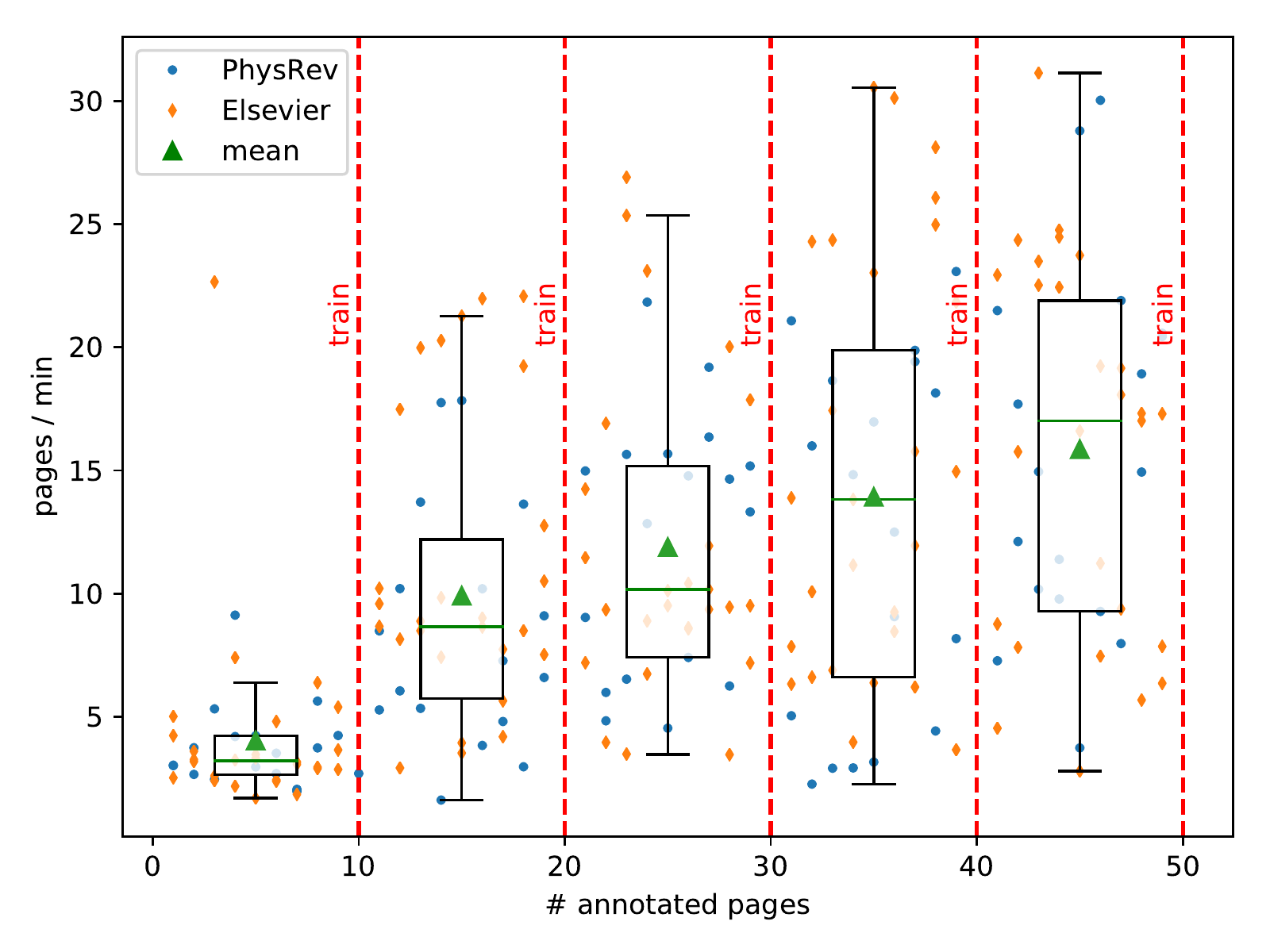}
\caption{\label{fig:SAAnnotationRate}The annotation rate of pages for two different collections (Physical Review B and Elsevier papers) as a function of the number of annotated pages. As one can observe, the mean annotation rate is increasing after each training (depicted by a vertical dashed red line). After the first training, the human annotator is presented a pre-annotated page, using the predictions from the latest model. As the predictions become better with increasing size of the ground-truth, less corrections need to be made and hence more pages can be annotated in similar time intervals.}
\end{figure}

The second purpose of the annotators is to visually inspect the quality of our machine learned models. The goal of the models is to emulate the action of the annotators, i.e. to assign a layout semantic label to each cell. Clearly, the result of a prediction for each page can therefore be displayed as if it were an annotated page. This allows the users to directly inspect the results of the models on unseen pages. A direct consequence of this inspection capability in the annotators is that the annotation task can be transformed easily into a correction task, i.e. the human annotators only need to correct the incorrectly predicted labels.
Of course, as the models become better over time, the number of corrections needed to be made become less and less. This allows us to significantly reduce the annotation time per document. Since annotations are typically created by professionals with a high hourly rate, the \textit{colouring} technique allowed us to significantly reduce the cost of ground-truth gathering.

In Figure~\ref{fig:SAAnnotation}, we show the annotation-rate in number-of-annotated-pages per minute. The vertical red lines indicate that a training was performed on the annotated pages, and a new, improved model is used from that point to predict the labels. Since the corrections become less and less, the rate of annotation goes up. It is needless to say that this inter-leaving of training models (based on annotated ground-truth) and annotation benefits directly from our platform approach, since each task (submitting page-annotations, training the model, applying the model for predicting the labels) comes down to an asynchronous call to a microservice. The accelerated annotation leads to a speed-up of a factor of 10 for ground-truth collection.

\subsection{Machine Learning: Training models \& Applying models}

In the CCS, there are essentially two types of machine-learning models. On the one hand, we have default models, which are designed to be layout independent. They take a raster image of the page to identify and locate basic objects, such as tables, figures, formulas, etc. On the other hand, we also support the training of custom, template-specific models, which are designed to specialize on a particular layout template and allow us to convert and extract the data out of documents with very high precision and recall. They will classify each cell in the page with regard to their layout semantic label.

\subsubsection{Metrics}
Before discussing the performance of the models, let us first define the precision and recall metrics used to evaluate the results.
The first observation is that the output of a machine learned model is exactly the same of what a human annotator would produce, i.e. it will assign a text cell a semantic label. The correctness of this label is what we aim to measure with the recall and precision metrics. The second observation is that we deal with a multi-class classification problem, i.e. we don't have only two labels, but many possible semantic labels, hence the performance result will be the average of the recall and precision for each label.

The recall (=$\mathcal{R}$) and precision (=$\mathcal{P}$) for a given label on a page is defined by the standard formulas
\begin{align}
\mathcal{R} = \frac{t_p}{t_p+f_p}, \quad  \mathcal{P} = \frac{t_p}{t_p+f_n},
\end{align}
where $t_p, f_p$ and $f_n$ represent respectively \textit{true positive}, \textit{false positive} and \textit{false negative} predicted labels.

\subsubsection{Default Models}

\begin{figure}[t]
\centering
\includegraphics[width=3.3in]{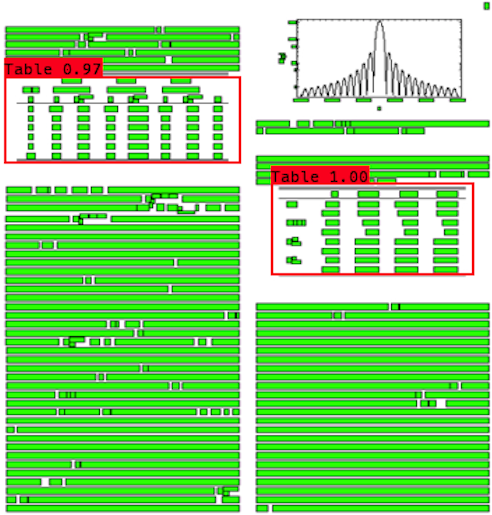}
\caption{\label{fig:SApredictedTable} A typical image of a parsed PDF page that is fed to the default models. In red, we show the detection of the tables combined with the confidence of the model. The results displayed here originate from the YOLOv2 model.}
\end{figure}

The aim of the default models is to identify specific, ubiquitous objects in documents. Examples of such objects are tables, figures with their captions, mathematical formulas, etc. Due to the high variability in both the document layout as well as in the representation of these objects, we need very robust object detection methods. Currently, the most robust methods for detecting objects are deep neural networks such as R-CNNs (and their derivatives Fast- and Faster-R-CNN)~\cite{GirshickDDM13:RCNN, Girshick:2015:FRCNN, NIPS2015_5638}, the YOLO architecture~\cite{Redmon2016YouOL, redmon2016yolo9000} and the SSD networks~\cite{Liu2016}. On our platform, we have the Faster-R-CNN~\cite{NIPS2015_5638} and the YOLOv2~\cite{ redmon2016yolo9000} networks available as individual microservices, both for training and predictions.

In this paper, we will focus only on the detection of table objects, but the same principles described in the following analysis are also applied for other type of objects.

The networks available on our platform have been trained on arXiv data\footnote{All the data is coming from the bulk data download \url{https://arxiv.org/help/bulk\_data\_s3}}. We have annotated 30000 PDF pages and know the location of at least one table on each page. From these 30000 pages, we have used 25000 pages as training data and kept the other 5000 pages for evaluation. Due to the large size of the dataset, we did not need to employ any data-augmentation technique, which is usually necessary for object-detection or image-classification algorithms.

We do not locate the table directly on the image of the original PDF page but rather on an  image representation of the parsed PDF page with cell boxes. The reasoning behind this is to reduce the variability between all input PDF pages as much as possible and thus increase the effectiveness of the deep neural networks. An example of such an image can be seen in Figure~\ref{fig:SApredictedTable}. The red bounding boxes around the tables are a result of the prediction using YOLOv2 and are absent in the image on which the model predicts. Note that the visualisation of the text cells visible in Figure~\ref{fig:SApredictedTable} does not include any text of the original document, but only its geometrical definition. This is important when one compares for example Asian documents with Japanese, Chinese or Korean characters versus European languages with the roman alphabet. We do not want the deep neural network to focus on the specific characters, but rather on the layout of the cells in the page.

\begin{table}[t]
\caption{\label{Table:TableDetection} Time-to-solution and performance results for the Faster RCNN and YOLOv2 models. The training of the models was done on 25000 PDF pages. The prediction (per page) and performance numbers (Recall=$\mathcal{R}$ and Precision=$\mathcal{P}$) were obtained on 5000 page images, where the prediction confidence cutoff was tuned to yield the maximum F1 score for each. All time-to-solution measurements for training were obtained on a POWER8 node with a single Pascal P100 GPU.}
  \begin{tabular}{lrrrr}
    \toprule
    \multirow{2}{*}{} &
      \multicolumn{2}{c}{Time to solution} &
      \multicolumn{2}{c}{Performance} \\
      & {Training} & {Prediction} & {$\mathcal{P}$} & {$\mathcal{R}$}\\
      \midrule
    Faster-RCNN & 72 hours & 4 sec & 0.97 & 0.98\\
    YOLOv2 & 9 hours & 0.1 sec & $0.99$ & $0.98$\\
    \bottomrule
  \end{tabular}
\end{table}

Let us now discuss both deep neural network training microservices on the platform. In Table~\ref{Table:TableDetection}, we show the time-to-solution for training and predicting a single page as well as the performance in terms of recall and precision. In the training phase, we ensure that both algorithms ran each 100 epochs, i.e. all 25000 page images were fed to the network 100 times. We observe that the \textit{out-of-the-box} Faster R-CNN from Tensorflow does not implement any batching during the training phase, while YOLOv2 batches 8 images at a time, thanks to an image resizing which is automatically applied. We believe that this is the main origin for the discrepancy of time-to-solution for the training phase. The same holds true for the prediction. Therefore, from the point of view of the platform, the YOLOv2 architecture seems better suited for deployment, as it allows to have a much higher throughput ($\approx$10 pages/sec/node).

For the performance analysis, let us outline one pre-processing stage which is needed before computing the metrics described previously. The object-detection networks predict a set of bounding boxes with a confidence level between 0 and 1. We use these bounding boxes to associate with each cell a label, which is in this particular case either \textit{Table} or \textit{Not-Table}, depending on whether they overlap with the predicted bounding box.
The corresponding recall and precision are then computed for this dual-class classification problem.
In order to do a fair comparison of the two networks, we optimise the precision and recall metrics with regard to the predicted confidence. For YOLOv2 we observe that the recall goes down and the precision goes up as the confidence is increased, obtaining a maximum F1 score of 98.7\% at a confidence level of $0.5$. The Faster R-CNN method is also performing quite well, but has slightly lower precision and recall numbers. We believe this originates from the selective search algorithm which is used to determine regions of interest. The images we feed it are not typical photographic images (made with a camera) but layout visualisations. The selective search algorithm in Faster R-CNN might not be optimal for such type of objects.

\subsubsection{Template specific Models}

The goal of template specific models is to obtain a better extraction quality by specializing the model on a specific template. This is necessary in many technical fields, where the accuracy of the extracted data is of paramount importance. Furthermore, many technical documents in a specific field typically appear in a certain template and it often makes sense to take advantage of this template to improve extraction quality.

For an algorithm to fit in the interactive platform design we identified a few key requirements. First, it is crucial that the model can generate good results with a limited set of pages. In practice this means the algorithm needs to perform well for 100-400 annotated pages, or the equivalent of a couple of man-hours for annotation. Second it must be robust against extreme imbalance of the labeled data. It is clear that cells of the label \textit{Title} will be much more uncommon than cells with the label of \textit{Text}. Last, the model needs to be very quick in training and predicting, since it will support the interactive annotation process.
\begin{table}[t]
\centering
\caption{\label{Table:PhysRevB} Performance results for the template specific model of the \textit{Physical Review B} journals. The confusion matrix highlights the huge imbalance between the number of text cells with different labels. The usage of ensemble machine learning methods allows to achieve a very high accuracy over all label types.}
\begin{tabular}{l|l|llllllll||l||}
\toprule
                  & & \multicolumn{6}{c}{predicted label} \\ \hline
\multirow{8}{*}{\rotatebox[origin=c]{90}{true label}} &  & \rotatebox[origin=l]{90}{Title} & \rotatebox[origin=l]{90}{Author} & \rotatebox[origin=l]{90}{Subtitle\:} & \rotatebox[origin=l]{90}{Text} & \rotatebox[origin=l]{90}{Picture} & \rotatebox[origin=l]{90}{Table} \\ \hline 
                   & Title             & 75 & 0    & 0     & 0         & 0      & 0 \\
                  &  Author         & 1  & 670 & 0     & 0          & 0      & 0 \\
                  &  Subtitle        & 0  & 0     & 325 & 0          & 0      & 0 \\
                  &  Text             & 1  & 17      & 0    & 56460 & 14      & 0 \\
                  &  Picture        & 0  & 0      & 0     & 4        & 4223 & 26 \\
                  &  Table           & 0  & 0     & 0      & 0        & 1       & 3418   \\ \hline \hline 
                  & Recall& 100 &  99.85  & 100      & 99.94        & 99.24     & 99.97 \\
                  & Precision& 97.40 &   97.52  & 100      & 99.99        & 99.64      & 99.24 \\
\bottomrule
\end{tabular}
\end{table}

For these reasons, we chose random forest~\cite{Breiman2001} as a machine learning algorithm for template specific models. Random forest algorithms are known to be trained fast and can produce very accurate results on limited, but relatively structured data. In our case, this structure originates of course from the template. Furthermore, random forest is an ensemble method, meaning that they learn on the distribution function of the features, and not individual data-elements.  As a consequence, they are typically more robust against imbalance of the labeled data, since the distribution functions are renormalised.

The random forest method is applied to each cell of the page based on a feature vector representing all of its properties. For example, the feature vector contains information as the page number, the size of the text cell, its position, as well as the distance from the neighbouring cells. Additionally to pure geometrical information we include the text style (normal, italic, or bold) and some text statistics, as the fraction of numeric characters. We then improve the obtained results by performing subsequent iterations with other random forest methods, which operate on an enlarged feature space including the previously predicted labels of the neighbourhood around the current cell.

It is important to realize that almost all of these features are purely geometrical. This allows us to apply exactly the same machine learning methods on both scanned and programmatic PDF documents.

In Table~\ref{Table:PhysRevB}, we illustrate the performance results of the models for a particular scientific journal, Physical Review B\footnote{\url{https://journals.aps.org/prb}}. We randomly chose 100 open-access papers and annotated 400 pages of them with 6 semantic labels. Tables~\ref{Table:PhysRevB} shows the confusion matrix between the true and the predicted labels as well as the derived recall and precision metrics for each label. We observe that the recall and precision numbers are excellent, with most of them above 99\%. This is not surprising, since we are building models that specialise for a particular template.

\begin{table}[t]
\centering
\caption{\label{Table:TemplatePerformance} Comparison for two different journal templates showing the aggregated precision and recall averaged over all labels. Each model has been independently trained on a dataset of 400 pages each. The results show that the ML algorithm proves to perform very well for the multiple document templates, simply by providing a different dataset to train on.}
  \begin{tabular}{lcc}
    \toprule
      Journal template & {$\mathcal{P}$} & {$\mathcal{R}$}\\
      \midrule
    Physical Review B & $98.96$ & $99.83$\\
    Elsevier & $99.46$ & $99.58$\\
    \bottomrule
  \end{tabular}
\end{table}

Moreover, the same ML algorithm proves to perform very well on different document templates, as is evident from the numbers shown in Table~\ref{Table:TemplatePerformance}, simply by providing it with different datasets to train on. The latter is the power of our platform: we can re-use the same machine-learning algorithm to generate different models solely based on the data gathered by the annotation on the platform. We do not need to define rules and heuristics or update code in order to deal with new types of documents. We only need to gather more data.

\begin{listing}[t]
\caption{\label{list:exampleoutput}Excerpt from the JSON output of the Corpus Conversion Service after conversion of this paper.}
\begin{minted}[breaklines,fontsize=\scriptsize]{json}
{
  "description": {
    "title": "Corpus Conversion Service: A machine learning platform to ingest documents at scale.",
    "abstract": "Over the past few decades, the amount of scientific articles [...]" ,
    "affiliations": "IBM Research Rueschlikon, Switzerland ",
    "authors": "Peter W J Staar, Michele Dolfi, Christoph Auer, Costas Bekas "
  },
  "main-text": [ {
      "prov": [ {
          "bbox": [ 52.304, 509.750, 168.099, 523.980 ],
          "page": 1
        } ],
      "type": "subtitle-level-1",
      "text": "1 INTRODUCTION"
    },
    {
      "prov": [ {
          "bbox": [ 52.304, 337.678, 286.067, 380.475],
          "page": 1
        } ],
      "type": "paragraph",
      "text": "It is estimated that [...] put these into context."
    },
    ...
  ],
  "tables": [ {...}, ... ],
  "images": [ {...}, ... ]
}
\end{minted}
\vspace{-0.2cm}
\end{listing}

\subsection{Assembly}
In this component, we build a structured data file in JSON or XML format, which contains all the text and objects (e.g. tables) from the original document, retaining the layout semantics. This structured data file is constructed by assembling all the cells from the parsed file in combination with their associated predicted (or human-annotated) layout semantic labels. It should be noted that no machine learning is used in this component. It is purely rule based and therefore completely deterministic.

The assembly phase is a two step process. First, one gathers all the cells with their associated layout semantic label and sorts them according to reading order. Then, the text of all cells that have the same label is contracted into a temporary document objects. Third, we build the internal structure of the temporary document objects, based on the information provided by the models. The latter is only applicable for internally structured objects, such as tables. An example of the generated JSON output is shown in Listing~\ref{list:exampleoutput}.


\section{\label{sec:Arch} Architecture and orchestration of Cloud based microservices}

In this section, we describe how the microservices in each of the components of the platform are deployed and orchestrated. Before discussing the technical details, we would like to point out our requirements for the architecture of the platform. These requirements are all related to scaling. Specifically, we would like the platform to scale with the number of documents, the number of users and last but not least the number of cloud based compute resources. In other words, we want a service that can ingest millions of documents, serve potentially thousands of users and scale its compute resources such that the time-to-solution is reasonable at all times for any operation. It is clear that the architecture of such a service is heavily influenced by these requirements.

\subsection{Platform layers}

\begin{figure}[!t]
\centering
\includegraphics[scale=0.28]{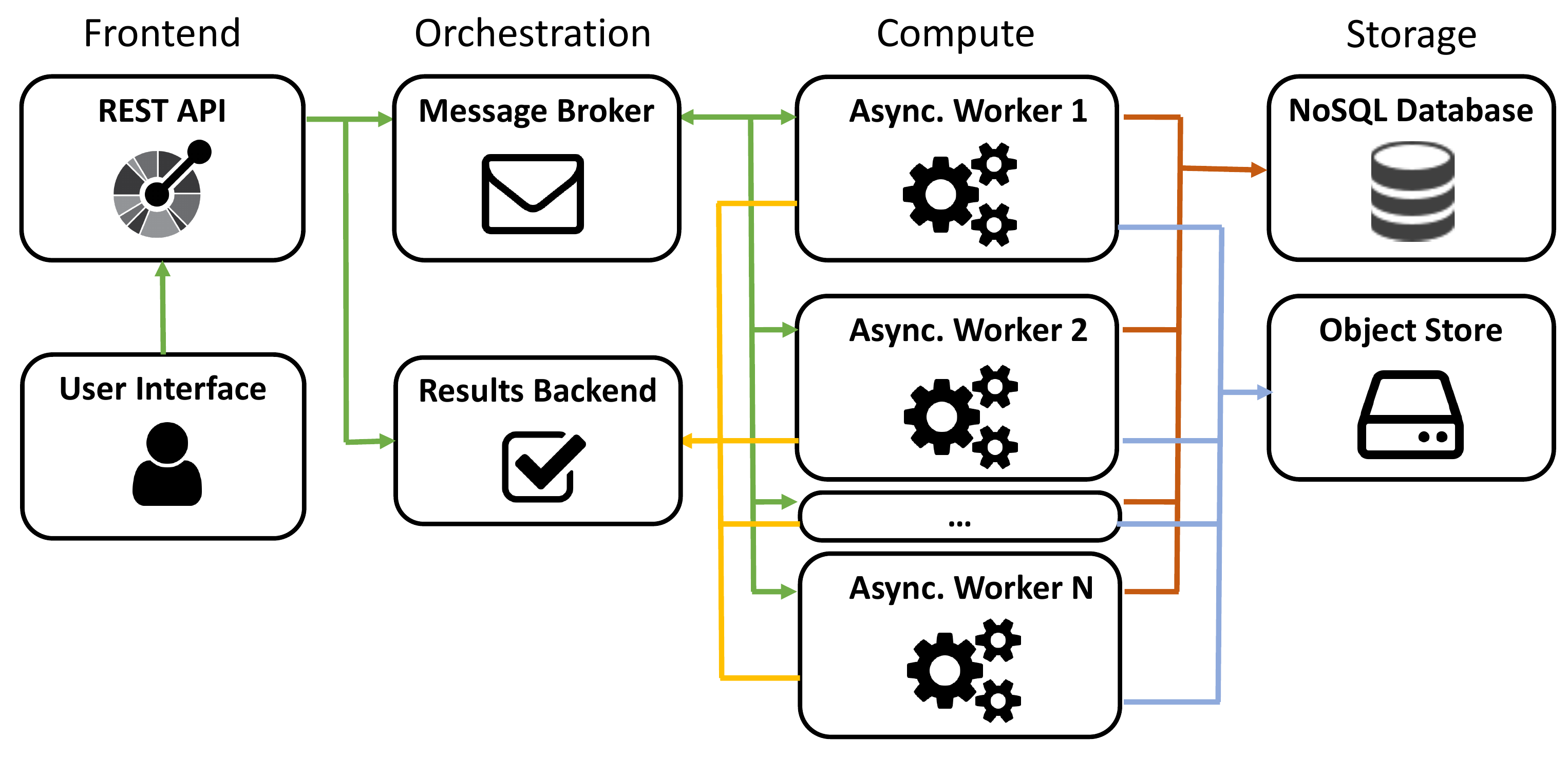}
\caption{\label{fig:SADesign} Diagram of the architecture of our platform. The architecture is composed from 4 layers: an interface layer with REST-API and frontend, an orchestration layer with a message broker and results backend, a compute layer consisting out of a variable number of asynchronous workers and finally a storage layer providing a NoSQL database and an object store. The NoSQL database stores the queryable meta-data of each file that is stored in the object store.}
\end{figure}

In Figure~\ref{fig:SApipeline}, we have shown a diagram of our pipeline on the platform to process documents. In Figure~\ref{fig:SADesign}, we show a sketch of its architecture. As one can observe, we have grouped the service into four layers. These layers are:

\begin{enumerate}
\item An interface layer which implements a REST-API and a user frontend: The user frontend is an AngularJS application build on top of the REST-API and implements the annotators for ground-truth gathering. The REST-API is built and documented using the OpenAPI specifications\footnote{\url{https://www.openapis.org/}} and is implemented in Python.
\item An orchestration layer that schedules the tasks for the microservices, stores their execution status and final result. The task scheduling is done with the Message Broker RabbitMQ\footnote{\url{https://www.rabbitmq.com/}}.
The results are stored in the in-memory data store Redis\footnote{\url{https://www.redis.io/}}.
In order to perform certain consecutive tasks (\eg parsing a PDF page with embedded scanned images requires first a parsing of the programmatic PDF page to extract the images and then an OCR service to extract the cells from these images) we can directly chain tasks, such that subsequent steps are only executed if the previous terminated successfully. This approach allows for a very robust, fault-tolerant service with very little downtime.
\item A compute layer that implements the microservices detailed in section~\ref{sec:Method}: Each of the workers in this layer executes the available microservices (e.g. parsing, training, predictions, assembly, etc). In order to scale with regard to resources, we have encapsulated each microservice into a distributed task queue using the Celery library\footnote{\url{http://www.celeryproject.org/}}. This allows us to dynamically scale the compute resources, since each worker can be spawned automatically on the cluster and register itself to the broker.
The workers are not only consumers of tasks, but may also produce new ones. This is the case for the requests operating on the whole corpus. Whenever possible we parallelise the compute-heavy operations at the page (or document) level.
\item A storage layer that stores all documents as well as the results from the microservices: The storage layer is composed out of two services: an object-store that stores all documents and processed stages (\eg the parsed PDF pages, trained models, etc) and a queryable NoSQL database that stores the metadata of each file in the object-store.
The object-store allows us to easily scale the storage with regard to the number of processed documents. However, it is not build to be queried efficiently, which is why we put a NoSQL database (in our case we use MongoDB\footnote{\url{https://www.mongodb.com/}}) on top to manage the storage and act as an access-layer.
\end{enumerate}

\begin{figure}[!t]
\centering
\includegraphics[width=3.5in]{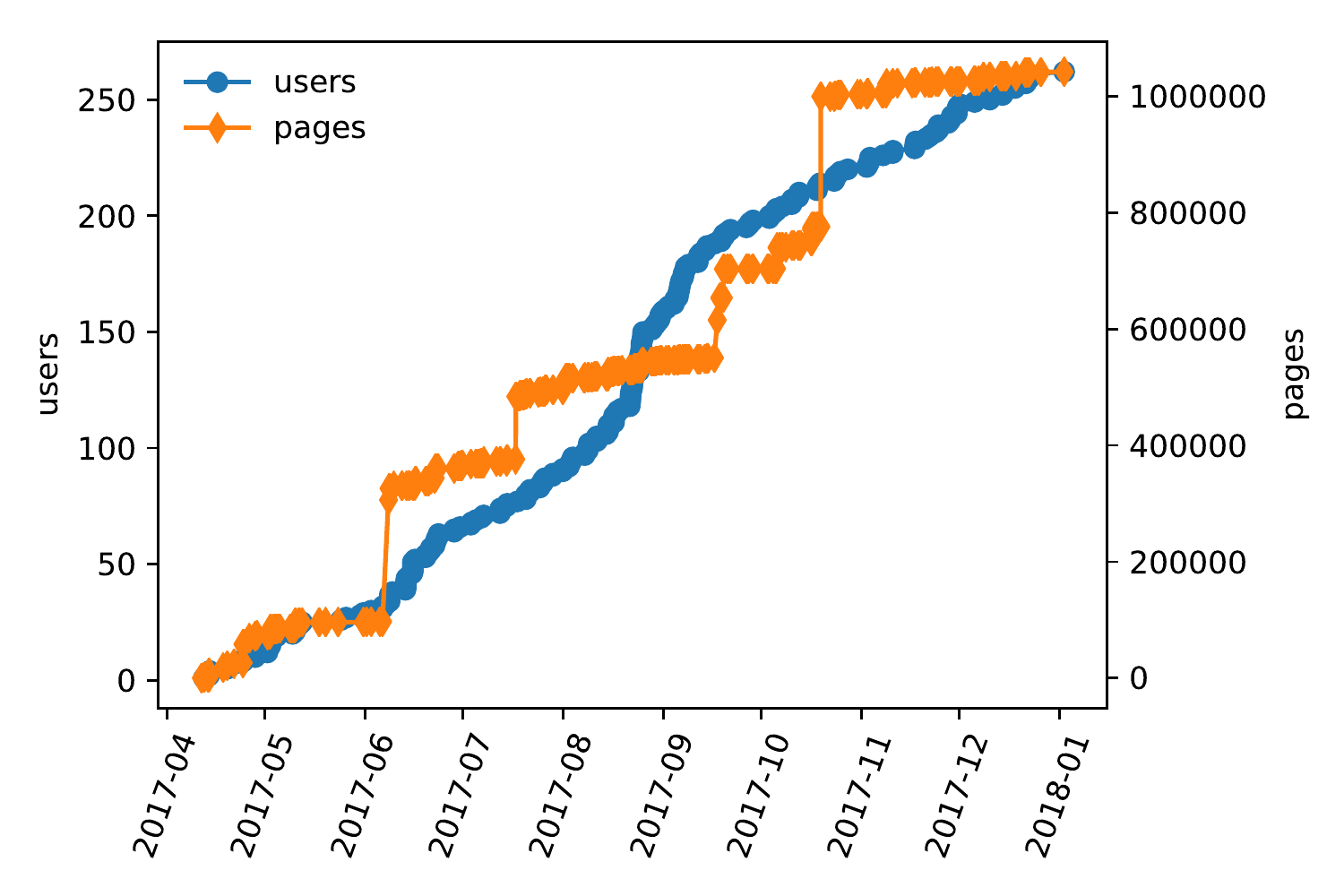}
\caption{\label{fig:ScalingUsersPages} Evolution of number of users and number of PDF pages on the platform. The jumps in the number of pages originates from big ingestions of documents performed by some users. This proves that the CCS platform is also able to accomodate these short burst of extreme activity.}
\end{figure}

By design, all the microservices in the compute layer are stateless, i.e. they don't manage any data, but only operate on it. This allows us to trust the additional stability and data safety concerns to the state-of-the-art tools that we have chosen, such as MongoDB, Redis and RabbitMQ. Being a cloud-based platform, our solution allows for these software assets to be detached from the main deployment and to be served by specialised vendors services which are certified to the latest industry requirements such as data-at-rest encryption, high availability, etc.

The choice of the services plays also a crucial role in addressing the scaling requirements for the platform. From the sketch (Fig.~\ref{fig:SADesign}), it is clear that the compute layer has a considerable amount of communication with these external services. During the development we evaluated multiple options and, \eg we had to replace some services because of inadequate performance or scaling bottlenecks. For example other result-backends didn't offer the auto-cleaning functionality offered by Redis and, before opting for a custom solution mixing MongoDB with an object storage, we evaluated other solutions as the GridFS storage, but it didn't fit to the constraints of typical cloud environments.

\subsection{Deployment}

Our platform is deployable on Kubernetes clusters\footnote{https://kubernetes.io/} available on many cloud providers or even on-premise installations, \eg using the IBM Cloud Private~\footnote{ibm.biz/privatecloud} distribution. Depending on the requirements, the storage services are launched inside the same cluster or linked to externally hosted endpoints.

The common parts of all deployments are the interface and the compute layer. The compute layer is designed for dynamically adapt the number of resources on the current load. For example, more parsing-microservice instances could be spawned when a large document is uploaded and they can automatically scaled down at the end of the task, such that the resources are free for other components, like training and assembling the processed documents.

The components running in the compute layer are further organized in different queues, such that we can control the fraction of resources allocated for each different component depending on their computational requirements. The parse component is indeed more demanding than the simple annotation components.

Currently, our main system operates on 5 Kubernetes nodes with 4 CPU cores and 8 GB of main memory each, and additionally one POWER 8 node with four GPUs is dedicated to the deep learning training and prediction tasks. Here, the flexible binding of microservices to specific nodes is a great advantage of the Kubernetes deployment. Moreover, 5 other virtual machines are employed to host the services in the orchestration and store layer.

\subsection{Scaling benchmarks}

\begin{figure}[!t]
\centering
\includegraphics[width=3.5in]{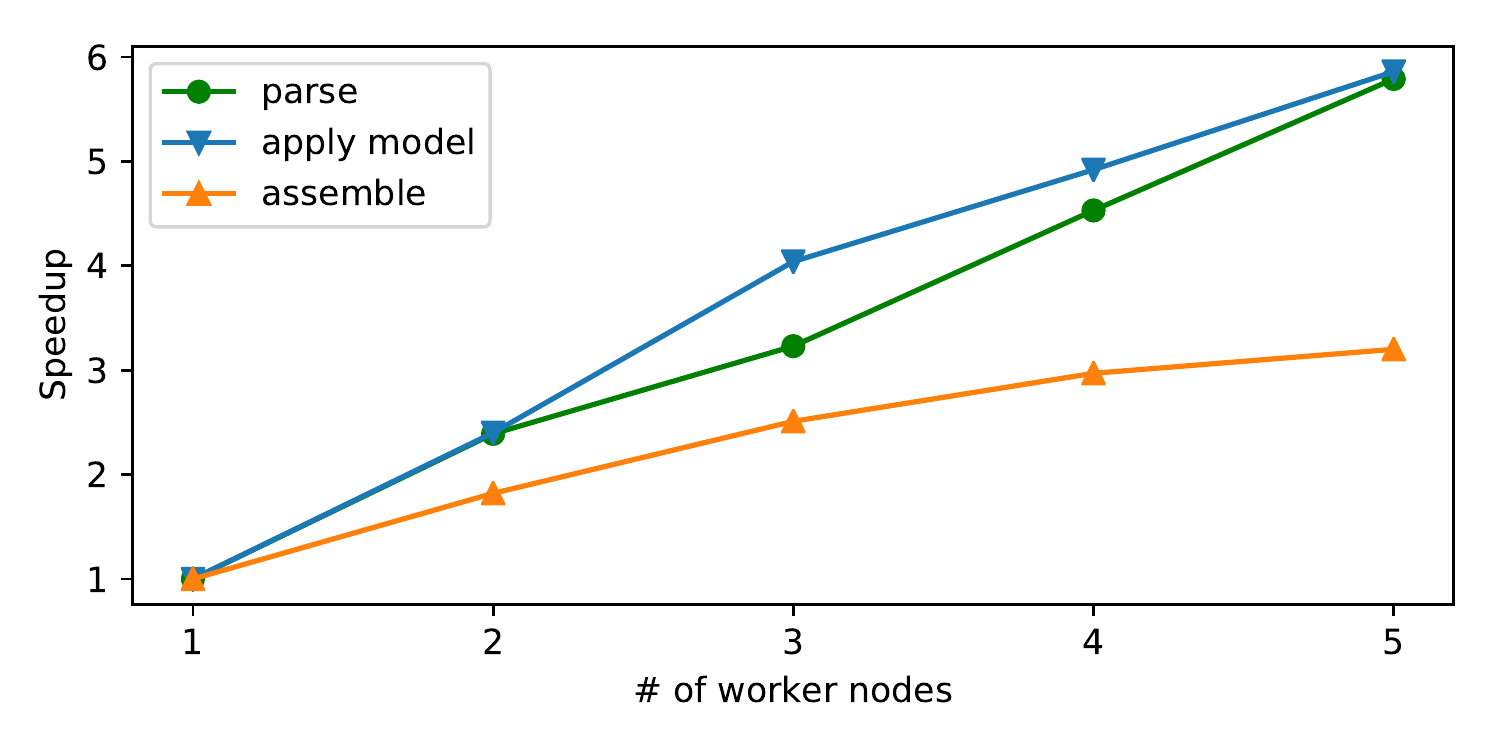}
\caption{\label{fig:ScalingResources} Speedup in the pipeline components as a function of the number of worker nodes (each with four cores, running four local worker processes).}
\end{figure}

Let us now discuss some scaling results on our platform. As we pointed out in the beginning of the section, our requirements for the platform were scaling with regard to the number of users, the number of processed documents and compute resources. In Figure \ref{fig:ScalingUsersPages}, we show the number of users and the number of processed PDF pages\footnote{We don't show the number of documents, since the number of pages in a document can range from 1 to well above 1000. Consequently, the number of pages is a more robust metric to measure the scaling with regard to the corpus size.} as a function of time. As one can see, the number of users and processed PDF pages has been increasing steadily over time since the launch of our service in April 2017. It is however interesting to see that there are sharp steps, indicating that some users have been uploading massive amounts of documents into the service in a very small amount of time. Due to our design, it was not a problem to accommodate these peaks and our service was able to handle these short burst of extreme activity.

In Figure \ref{fig:ScalingResources}, we show the scaling of the three main pipeline microservices (i.e. the parsing of PDF documents, applying machine learned models and conversion of documents to JSON) on the platform with regard to compute resources. We show this scaling by displaying the speedup versus the number of worker nodes available. Here, we chose to have four workers serving each pipeline microservice, since each worker is running on a node with four cores. As one can observe, the speedup in the parse and ML apply tasks scales linearly with the the number of workers, and thus the nodes. Notably, we can even observe a slightly better-than-linear speedup, which appears due to bandwidth constraints on the baseline with one worker. The speedup on the assemble tasks, in comparison, flattens off sooner, as this task can only be parallelised on the document and not on the page level.
The variability in the length of documents is reflected in a load imbalance between the worker nodes, however this averages out with sufficiently large corpus sizes. Consequently, we are able to scale the compute resources in order to keep the time-to-solution constant for any job-size.

\section{\label{sec:Conclusion}Conclusion}

We have presented a scalable, cloud based platform, which can ingest, parse and annotate documents, and particularly, train \& apply advanced machine learning models in order to extract the content of the ingested documents and convert it into a structured data representation.

The fundamental design choices in our solution have proven to enable scaling in three elementary ways. First, it can service multiple users concurrently. Second, it can ingest, parse and apply machine learned models on many documents at the same time. Third, it can scale its compute resources for different tasks on the platform according to their respective load so the conversion of documents on the platform is at all times bounded in time, given enough resources.

In the future, we plan to extend the platform in two major areas. First, we would like to extend the number  of microservices, especially with regard to image understanding. The number of types of images is enormous (e.g. line \& scatterplot, histograms, pie-charts, geographic maps, etc). The goal here would be to extract the data out of these individual type of images after a successful identification with an image-classifier. Second, we would like to improve the quality and performance of our default models. We strongly believe that the results can be greatly improved since the neural networks we currently use are optimised for photographic images, and not images of parsed document pages (as is shown in Figure~\ref{fig:SApredictedTable}).
To leverage this growing use of deep learning models, we will additionally introduce specialised data-parallelism in order to speed up the training and provide interactive user-customisation capabilities.

\begin{acks}
  The authors would like to thank Roxana Istrate and Matthieu Mottet for their
  contribution to the development of the CCS system.

  This work was supported by the \grantnum[http://nccr-marvel.ch]{}{NCCR MARVEL},
  funded by the \grantsponsor{}{Swiss National Science Foundation}{}.
  MD was supported by the FORCE project, funded by \grantsponsor{NMBP-23-2016}{Horizon 2020}{} under NMBP-23-2016 call
  with Grant agreement number \grantnum[http://the-force-project.eu]{NMBP-23-2016}{721027}.

\end{acks}

\bibliographystyle{ACM-Reference-Format}
\bibliography{acmart}

\end{document}